\documentclass{article}
\usepackage{spconf,amsmath,graphicx}
\usepackage{colortbl}
\usepackage{url}
\usepackage{color}
\usepackage{subfigure}
\usepackage[table]{xcolor}
\usepackage{booktabs}
\usepackage{multirow}

\definecolor{darkgreen}{rgb}{0.7176, 0.7843, 0.7176}
\definecolor{lightgreen}{rgb}{0.9529, 0.9569, 0.9529}

\colorlet{tableheadcolor}{darkgreen} 
\newcommand{\headcol}{\rowcolor{tableheadcolor}} %
\colorlet{tablerowcolor}{lightgreen} 
\newcommand{\rowcol}{\rowcolor{tablerowcolor}} 
\newcommand{\topline}{\arrayrulecolor{black}\specialrule{0.1em}{\abovetopsep}{0pt}%
	\arrayrulecolor{tableheadcolor}\specialrule{\belowrulesep}{0pt}{0pt}%
	\arrayrulecolor{black}}
\newcommand{\midline}{\arrayrulecolor{tableheadcolor}\specialrule{\aboverulesep}{0pt}{0pt}%
	\arrayrulecolor{black}\specialrule{\lightrulewidth}{0pt}{0pt}%
	\arrayrulecolor{white}\specialrule{\belowrulesep}{0pt}{0pt}%
	\arrayrulecolor{black}\hline}

\newcommand{\bottomlinec}{\arrayrulecolor{tablerowcolor}\specialrule{\aboverulesep}{0pt}{0pt}%
	\arrayrulecolor{black}\specialrule{\heavyrulewidth}{0pt}{\belowbottomsep}}%

\title{Provenance Filtering for Multimedia Phylogeny}

\name{A. Pinto$^{1,2}$, D. Moreira$^1$, A. Bharati$^1$, J. Brogan$^1$,\textit{K. Bowyer$^1$, P. Flynn$^1$, W. Scheirer$^1$ and A. Rocha$^{1,2}$}\thanks{This material is based on research sponsored by DARPA and Air Force Research Laboratory (AFRL) under agreement number FA8750-16-2-0173. Hardware support was generously provided by the NVIDIA Corporation. We also thank the financial support of FAPESP (Grant \#2015/19222-9),  CAPES (DeepEyes Grant) and CNPq (Grant \#304472/2015-8).}}

\address{$^1$Department of Computer Science and Engineering, Univ. of Notre Dame, IN, U.S.A.\\$^2$ Institute of Computing, Univ. of Campinas, SP, Brazil}
	
\begin{document}
\onecolumn

\noindent \copyright 2017 IEEE. Personal use of this material is permitted. Permission from IEEE must be obtained for all other uses, in any current or future media, including reprinting/republishing this material for advertising or promotional purposes, creating new collective works, for resale or redistribution to servers or lists, or reuse of any copyrighted component of this work in other works.
\\

\noindent Pre-print of article that will appear in IEEE International Conference on Image Processing (ICIP).
\\


\twocolumn
\newpage
\maketitle
\ninept

\begin{abstract}
Departing from traditional digital forensics modeling, which seeks to analyze single objects in isolation, multimedia phylogeny analyzes the evolutionary processes that influence digital objects and collections over time. One of its integral pieces is provenance filtering, which consists of searching a potentially large pool of objects for the most related ones with respect to  a given query, in terms of possible ancestors (donors or contributors) and descendants. In this paper, we propose a two-tiered provenance filtering approach to find all the potential images that might have contributed to the creation process of a given query $q$. In our solution, the first (coarse) tier aims to find the most likely ``host'' images --- the major donor or background --- contributing to a composite/doctored image. The search is then refined in the second tier, in which we search for more specific (potentially small) parts of the query that might have been extracted from other images and spliced into the query image. Experimental results with a dataset containing more than a million images show that the two-tiered solution underpinned by the context of the query is highly useful for solving this difficult task. 
\end{abstract}


\section{Introduction and Related Work}
\label{sec:intro}
\begin{figure}[htb]
	\begin{center}
	\subfigure[Semantically-similar \& near-duplicate images.\label{provenance_nndr}]{
		\includegraphics[width=0.67\columnwidth]{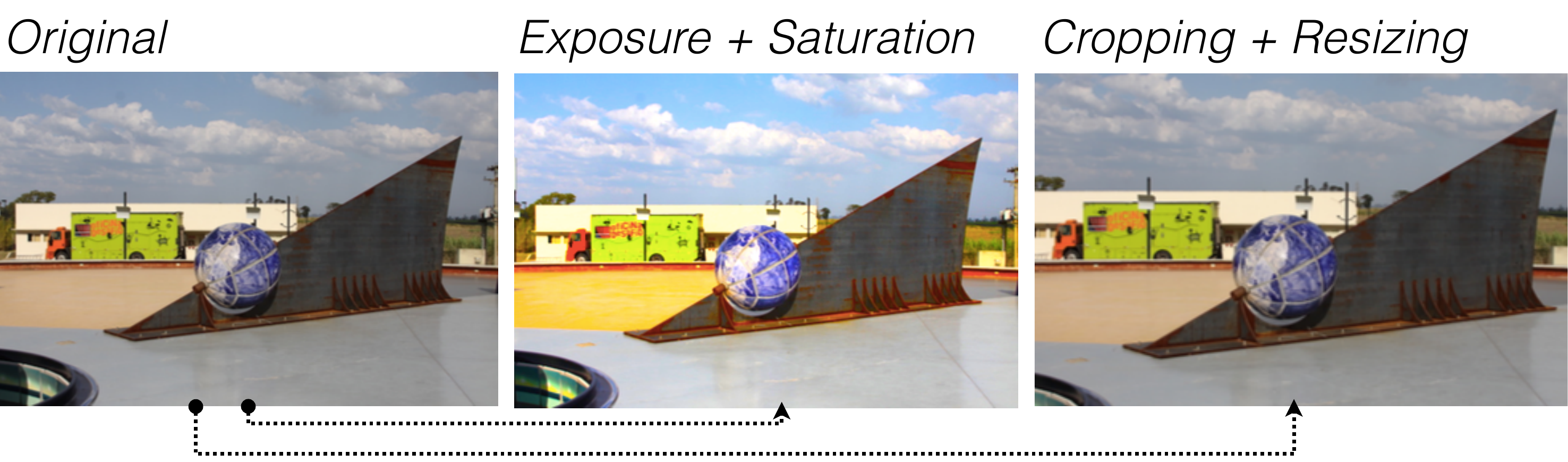}
	}
	\subfigure[Multiple parenting multimedia phylogeny setup with an image composition and its several ancestors (donors)\label{provenance_mpp}.]{
		\includegraphics[width=0.67\columnwidth]{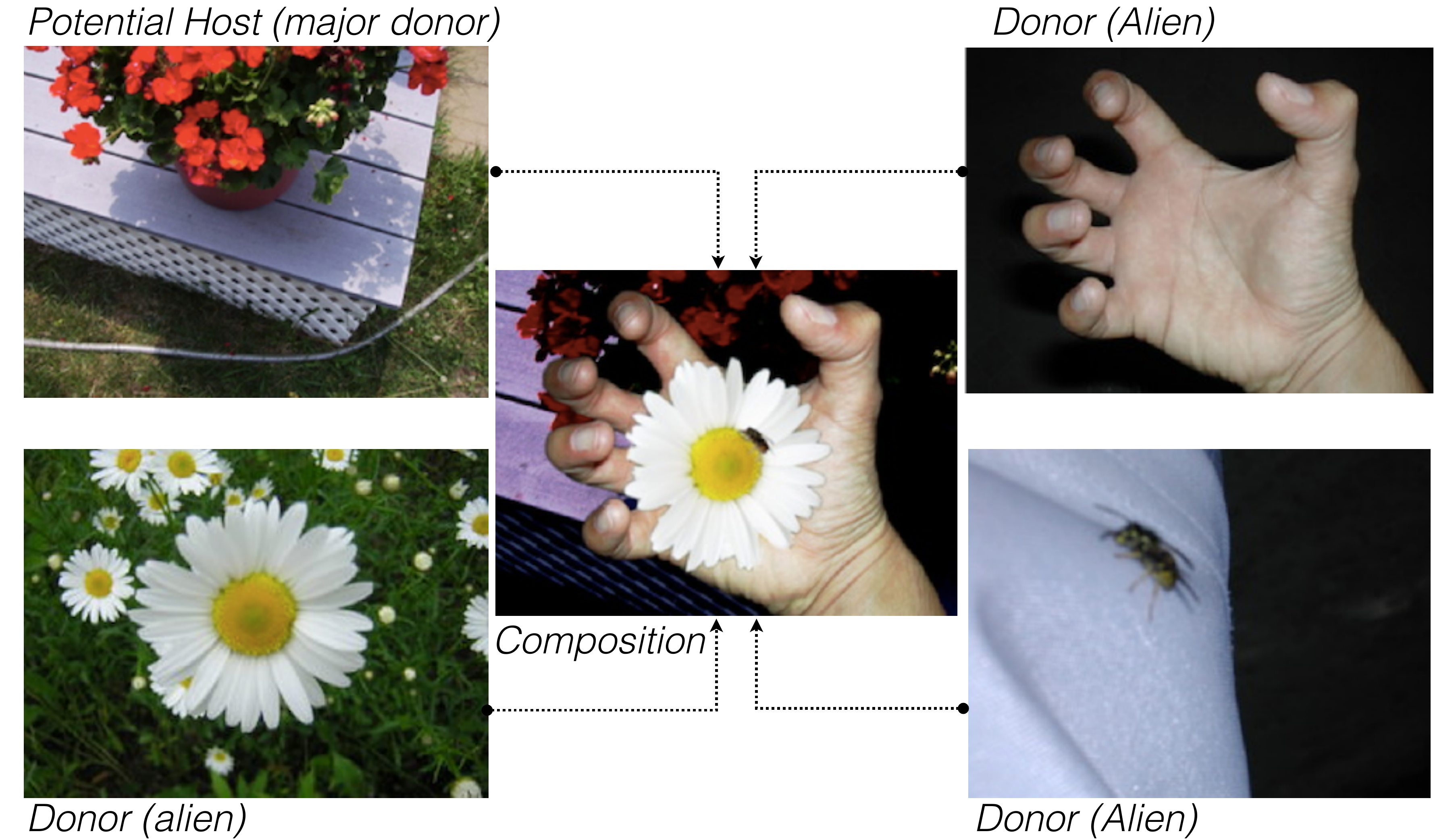}
	}
	\caption{Contrasting multimedia phylogeny applied to near duplicate images (a) and image composites with several donors (b). While the former focuses on finding relationships among images that have similar overall context, the latter aims at finding the genealogy of an asset, including all possible near duplicates of the composition itself and of its donors. Example in (a) from~\cite{Dias_2013}; example in (b) from the NIST Nimble 2016 dataset~\cite{Nimble_2016}.}
	\label{fig:fig1}
	\end{center}
\end{figure}


Rather than focusing on checking the integrity of a single multimedia object (as it used to be with most of the proposed methods from the early 2000s until recently), some researchers in digital forensics are now seeking to leverage all possible information associated to a pool of objects, analyzing their space and time relationships. Such recent efforts are made possible by a research field known as Multimedia Phylogeny~\cite{Dias_2012,Dias_2013} --- a relatively new discipline that studies the evolutionary processes that influence multimedia objects and collections, as well as the relationship among transformed versions of an object, looking for causal and ancestry relationships, the types of transformations, and the order in which they were applied to objects. 

Such new developments are necessary in order to adapt forensics methods to a rapidly evolving society. The increasingly frequent occurrence of image and video compositions on the Internet and social media render the applications of phylogeny very useful in practical scenarios such as content tracking, forensics and copyright enforcement~\cite{Dias_2012,Dias_2013}. Within this new reality, forensics analysts are interested not only in determining if a digital object is fake or real but also in pinpointing who created it, what happened, when and how (genealogy) an asset was created. This process might be of significant importance in the era of post-truth~\cite{Keyes_2004,Mahler_2016,Schulten_2017} for determining how a composition was crafted, what parts went into creating the composite, and whether there was re-staging, re-purposing or an overall change of semantics~\cite{Rocha:CSUR:2011}. 

Nonetheless, before analyzing a pool of objects looking for possible kinship relationships, we need to be able to comb through large quantities of data looking for the very pieces potentially associated with a given query $q$. This task needs to be performed prior to subsequent multimedia phylogeny steps --- namely the pairwise image dissimilarity calculations and the phylogenetic graph analysis and construction --- and it is referred to herein as \emph{provenance filtering}. 

Most of the work thus far in multimedia phylogeny has overlooked the provenance filtering task, considering it to be a reasonably well solved problem~\cite{Dias_2012,Dias_2013}. The rationale behind that assumption was that most phylogeny works focused on finding the evolutionary processes associated with near-duplicate~\cite{Dias_2012} and semantically-similar images~\cite{Dias_2013}. In both setups, original images may undergo transformations over time but cannot have their overall semantics changed. 
When we consider forged and composite images, we bring new elements to the table. In this case, we now have the appearance of multiple parenting phylogeny~\cite{Oliveira_2016}, a setup in which an image might be the composite result of several other images, each with its own evolutionary chain of modifications. The composite image itself might also have its own chain of descendants and so on. Fig.~\ref{provenance_nndr} shows an example of semantically-similar images in which an original image might undergo several transformations and generate offspring. Each child can also generate others. However, the transformations tend to keep the overall meaning of the scene. In turn, as we see in Fig.~\ref{provenance_mpp}, an image in a multiple parenting setup might be the result of combining several others, each of which having its own chain of ancestors and descendants. 

Near-duplicate detection (NDD) methods~\cite{Ke_2004,Zhou:MM:2010,Tang:ICIP:2015,Yuan:ICIP:2015, Zeng:ICIP:2016} work properly for the task of finding semantically-similar images (Fig.~\ref{provenance_nndr}), upon which phylogeny graph construction algorithms could operate later on. However, NDD methods might fail in the presence of multiple donors (Fig.\ref{provenance_mpp}) given that the context and meaning of each donor is too diverse to be represented and captured by current methods. Moreover, each donor might undergo several transformations in the composition creation process including color, geometric, and affine operations. For those cases, even partial near-duplicate detection methods could fail~\cite{Dong:ICMR:2012}. Likewise, traditional content-based image retrieval (CBIR) methods~\cite{Datta_2008} would not work directly either as they often aim to determine the overall meaning of the scene and its generalization to provide the user with similar images respecting the principles of novelty and diversity~\cite{Deselaers_2009}. 

While related work for multimedia phylogeny abounds, prior work on  provenance filtering is almost non-existent. In terms of phylogeny, Dias et al.~\cite{Dias_2012} presented a minimum spanning tree-based algorithm to find a directed graph that represented the phylogeny tree of a group of near-duplicate images. This work was extended to deal with images from multiple cameras and their near duplicates~\cite{Dias_2013}. Other media have also been considered such as videos~\cite{Dias_2011,Lameri_2014}, audio~\cite{Nucci_2013} and text~\cite{Andrews_2012}. Oliveira~et~al.~\cite{Oliveira_2016} extended the image phylogeny formulation to deal with multiple donors and descendants simultaneously more aligned with the context of this paper. However, their work assumes the candidate images are known a priori.  

Important advances have been made on finding ancestral relationships between pairs of images; nevertheless, the performance of such algorithms is significantly degraded if a good set of potentially related images is not found beforehand. In this vein, we extend upon image representation and indexing techniques (common in NDD and CBIR areas) to deal with provenance filtering for multiple donor and composite images. Our technique comprises two stages: in the first, we query an image collection for the most likely donors that might have contributed to the creation of the query, if it is a composite. This is done following a traditional CBIR pipeline, which involves image representation through appropriate features and the adoption of a subsequent indexing mechanism (more details in Sec.~\ref{sec:method}). The top retrieved results are then analyzed and compared to the query using scale and rotation-invariant points of interest~\cite{Bay:CVIU:2008}, nearest neighbor distance ratio policy~\cite{Lowe_1999}, and geometric alignment~\cite{Zitova_2003}. After finding the best possible match to the query, we use that image along with the query to calculate a contextual mask to serve as an activation of possible regions that are different between them. Such regions are candidate regions for possible donors. We then proceed with the second stage of the search, querying the collection for images that are similar to the selected regions of interest in the query as pointed out by the contextual mask. Ultimately, we aggregate the different rankings to create a final ranked list of images related to the query in terms of possible donors contributing to its creation process and thus closing the loop for provenance filtering.

The contributions of this work are (i) the exploration of different querying and indexing techniques for the new problem of provenance filtering; (ii) the incorporation of provenance context to single out possible candidate regions related to donors in the creation processo of a query; and (iii) the study of the efficiency and effectiveness tradeoffs involved in the provenance filtering task while dealing with very large collections of images. 


\section{Proposed Method}
\label{sec:method}
\label{sec:proposedmethod}

In this section, we present the proposed approach to provenance filtering. Given a query $q$, such as the image in the center of Fig.~\ref{provenance_mpp}, the objective is to search a collection of images $\mathcal{C}$ for all potential donors $r_i$ contributing to the creation of $q$, including possible near duplicates $r_{ij}$ of $r_i$. Near duplicates of $q$ are also of interest as they would be important for tracing the offspring of $q$ over time. 

\begin{figure}[t]
	\begin{center}
	\includegraphics[width=0.95\columnwidth]{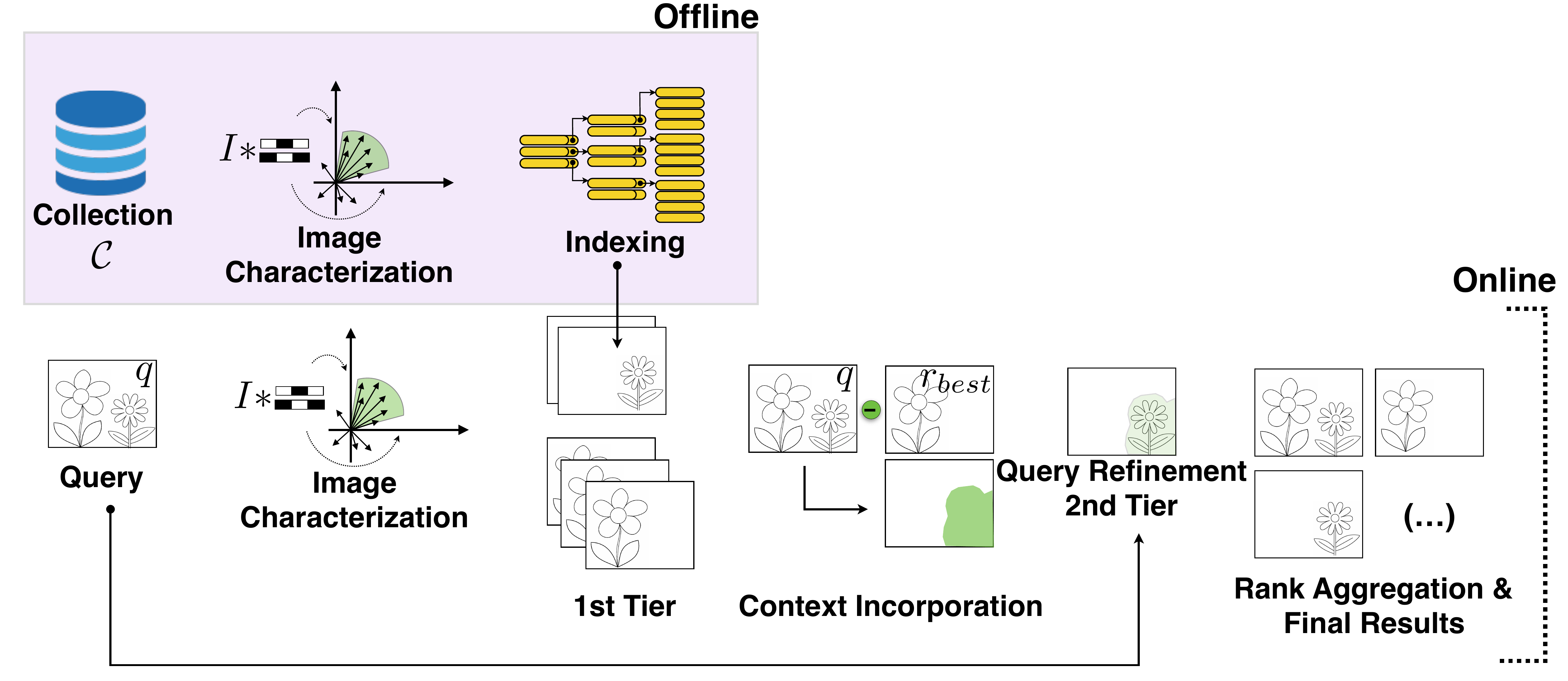}
	\caption{Method's pipeline. After retrieving related images, we compare the best result with $q$, incorporate the search's context and perform a second search to refine the list of possible donors.\label{fig:fig2}}	
	\end{center}
\end{figure}

Our approach to this problem involves two stages (c.f., Fig.~\ref{fig:fig2}). In the first stage, we design a fast image retrieval solution to recover the (likely) donor images, with high precision. We then exploit the context of the results to find the best match $r_{best}$ (respecting geometric constraints) with respect to $q$ and refine the donor list. Regions that are different between $q$ and its top-related image $r_{best}$ are of interest as they show regions that might have been incorporated into $q$ by combining pieces of different images in $\mathcal{C}$. Leveraging the contextual mask, the second stage of the search examines $\mathcal{C}$ a second time, focusing on finding potential localized donors. 

In the example of Fig.~\ref{provenance_mpp}, when querying the collection for potential donors (first tier/stage), we would likely retrieve the image with the table, flower and their background or the hand (as both are major contributors to the composite $q$). Calculating the contextual mask gives the region of the hand as a potential donor spliced from another source image(s). Therefore, when performing the second search, we look for images similar to that region, which would result in the donor for the hand as well as the other pieces. This process can be repeated a number of times if necessary. The different retrieved lists of results might be combined through rank aggregation techniques based on the confidence of the retrieved results. 

\subsection{Image Characterization}

The first step of our approach needs to represent each image in a robust manner so as to allow us retrieve partially related images in a large collection. In this context, using bags of words~\cite{Datta_2008} or deep learning techniques~\cite{Goodfellow_2016} would likely fail as they would be good for retrieving similar images in general but would not capture possible transformed donors, especially the small or heavily processed ones. In addition, a deep learning solution would require large image collections spanning different forgeries for a proper training and, in forensics, such collections are simply not available. In face of these limitations, we opted to represent each image using points of interest robust to image transformations, as forgeries often employ such transformations for more photorealistic montages. For that, we rely upon Speeded-Up Robust Features (SURF)~\cite{Bay:CVIU:2008}. We represent an image with about 2000 keypoints for small-scale experiments and with about 500 keypoints for large-scale ones. 

\subsection{Indexing Structure}
Given a query image $q$ and a collection of images $\mathcal{C}$ for searching, we need to represent the images in $\mathcal{C}$ in a very compact fashion so as to allow fast querying. For that, we use an indexing algorithm for finding nearest neighbors of $q$, in terms of their representative keypoints. More specifically, after extracting the points of interest for all images in $\mathcal{C}$, we need to find the $k$-nearest points to each keypoint in $q$. We further perform majority voting to infer the similarity between the query image $q$ and each image in $\mathcal{C}$ based on the nearest keypoints retrieved from the gallery.

As the number of points of interest extracted from $\mathcal{C}$ might reach hundreds of millions, the comparison between the $q$ and all images in $\mathcal{C}$ using brute-force search is impracticable. Therefore, we investigated some algorithms for $\epsilon$-approximated nearest neighbors, adequate for large-scale searches. According to Arya~\cite{Arya_1998}, an approximate search can be achieved by considering (1 + $\epsilon$)-approximate nearest neighbors for which $dist(k,l) \leq (1+\epsilon)dist(p,l)$ such that $p$ is the true nearest neighbor for $l$. Nonetheless, these solutions might lose effectiveness depending on the heuristic adopted to speed up the search. For this reason, here we compare four indexing approaches in terms of runtime, memory footprint and quality of the search: KD-Trees and KD-Forests~\cite{Bentley:COMMUN:1975}, Hierarchical Clustering~\cite{Steinbach:KDD:2000}, and Product Quantization~\cite{Jegou:TPAMI:2011}. 

\subsection{Context Incorporation and Ranking Aggregation}
To retrieve the donor images with high recall rates, we propose a query refinement process, referred to as context incorporation, in that we use the ranking result obtained in a first tier to reformulate the query so that small objects used to compose the spliced image can be be retrieved more accurately. First, we need to make sure the query is well represented in terms of describing keypoints. The overrepresentation of the query $q$ aims at guaranteeing we sample basically all of its regions, including the background. Although SURF descriptors are robust to describe objects in general in a scene, this approach most likely will fail in finding interest points inside very small objects, mainly when such objects are put in a complex background. To overcome this problem, we perform a query refinement by computing the intersection between $q$ and the best-matching retrieved image (most likely the host / background donor). This leads to a new query image containing just the information about the objects added in the host image. Our second search stage consists of querying the collection using the keypoints falling within the selected regions of interest. We combine the different ranked lists using the confidence of the retrieved images (number of votes and keypoints matched).

\subsection{Finding the Contextual Mask}
To find the contextual mask, we perform an image registration between $q$ and the top-match image $r_{best}$ in the ranked list obtained in the first tier of search. We match SURF features extracted from both images, select the $25$ best-matching keypoints and calculate the distance between the two images using the selected pairs of matches. We then calculate the geometrical transformation present in $r_{best}$ with respect to $q$ via image homography. Next, we compute the mask that indicates the candidate regions in which we might have spliced objects. We generate this mask by computing the difference between geometrically aligned images, followed by an opening operation with a $5\times5$-structuring element and a $5\times5$-kernel median filter to reduce the residual noise present in the mask. We also perform color quantization to 32-bits before computing the difference between the two images to reduce the presence of noise in the mask. 

\begin{figure}[t]
	\begin{center}
	\includegraphics[width=0.6\linewidth]{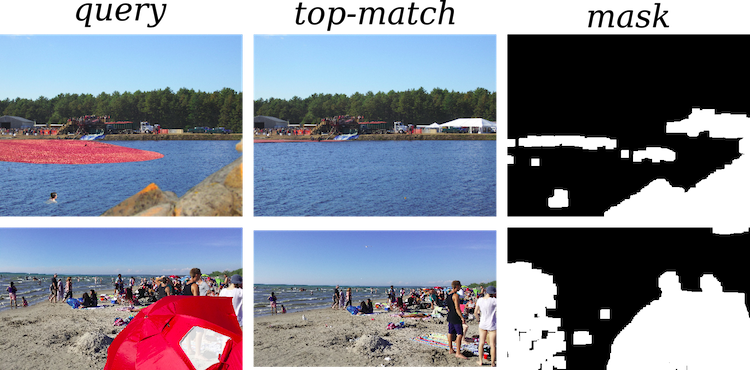}
	\caption{Example of a query, its top-related donor and the contextual mask. In the top row, the contextual mask captures the added rocks, person, bird and red-dirty region. In turn, the mask in the second row captures the added umbrella, content-smoothed sand on the left and the deleted white bird. 
	\label{fig:contextual:mask}}
	\end{center}
\end{figure}

There are some extreme cases for this approach that are worth discussing. First, when the top retrieved image does not have anything in common with $q$, the calculated mask should be null. In this case, there should be no search in the second tier. In turn, when $q$ itself is not a composite, the top retrieved image might be non-related at all (case one above) or a near-duplicate of $q$, in which case the mask is virtually identical to $q$. In the latter case, the search in the second tier should result in basically the same images retrieved in the first tier. Fig.~\ref{fig:contextual:mask} depicts examples of a query $q$, its top result $r_1$ and the calculated contextual masks. 

\section{Experiments and Results}
\label{sec:experiments}
In this section, we present and discuss the experimental results we performed to validate the proposed method. We report the quality of the results in terms of Recall@k that measures the fraction of correct images at the top-$k$ retrieved results. The source code of all proposed methods are freely available\footnote{The source code is freely available on \url{https://gitlab.com/notredame-provenance/filtering}}.

\vspace*{0.1cm}
\noindent
\textbf{Datasets.} 
We adopt the Nimble Challenge 2016 (NC2016) and 2017 (NC2017) datasets, provided by the National Institute of Standards and Technology (NIST)~\cite{Nimble_2016}, which focus on forensics, provenance filtering and phylogeny tasks. These datasets comprise a query set containing different kinds of manipulated images (e.g., copy-move and compositions), and a gallery set containing the source images used to produce the queries. The datasets also comprise distractor images. The probe sets of NC2016 and NC2017 datasets contain $288$ and $16$ composite images, respectively. The gallery sets contain $874$ and $10446$ images, respectively. We also embed the datasets within one million images (distractors) provided by RankOne Inc.\footnote{\url{http://medifor.rankone.io/}}, as recommended by NIST for evaluating scalability. 

\vspace*{0.1cm}
\noindent
\textbf{Indexing Method.} We now analyze (see Table~\ref{tab:indexing_method}) different indexing approaches for NC2017 and NC2017+World1M in terms of memory footprint and efficiency (results for NC2016 are similar) considering an Intel(R) Xeon(R), CPU E5-2620 v3 @2.40GHz, 24 cores and 512GB of RAM. Although PQ is more efficient in terms of storage for a small scale, it does not scale for World1M. The clustering in HCAL prevented it from scaling for 1M images. More work involving approximate clustering and sampling would be necessary in this case. KD-Tree shows a good storage and efficiency tradeoff. 
\begin{figure}[t]
	\begin{center}
	\includegraphics[width=0.48\linewidth]{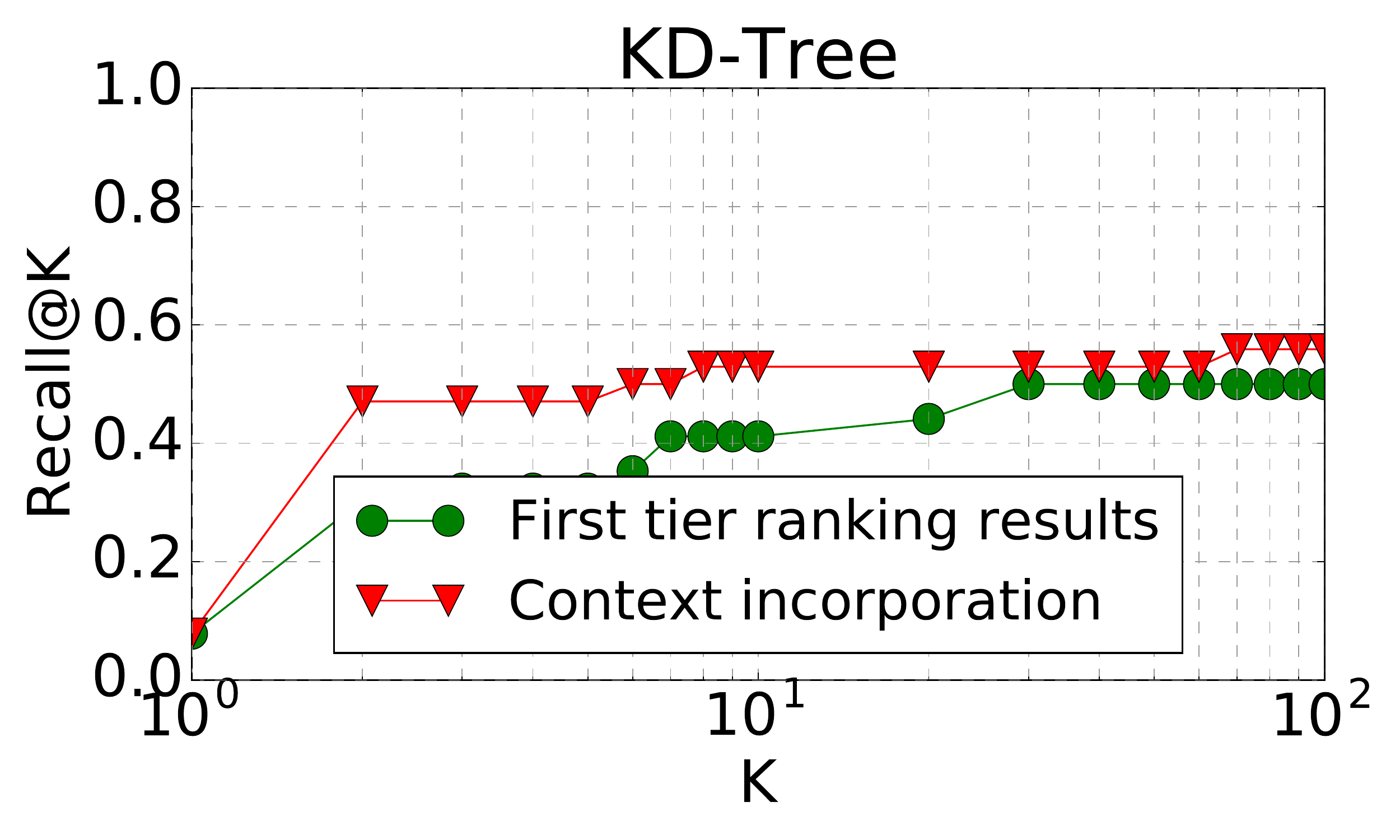}
	\includegraphics[width=0.48\linewidth]{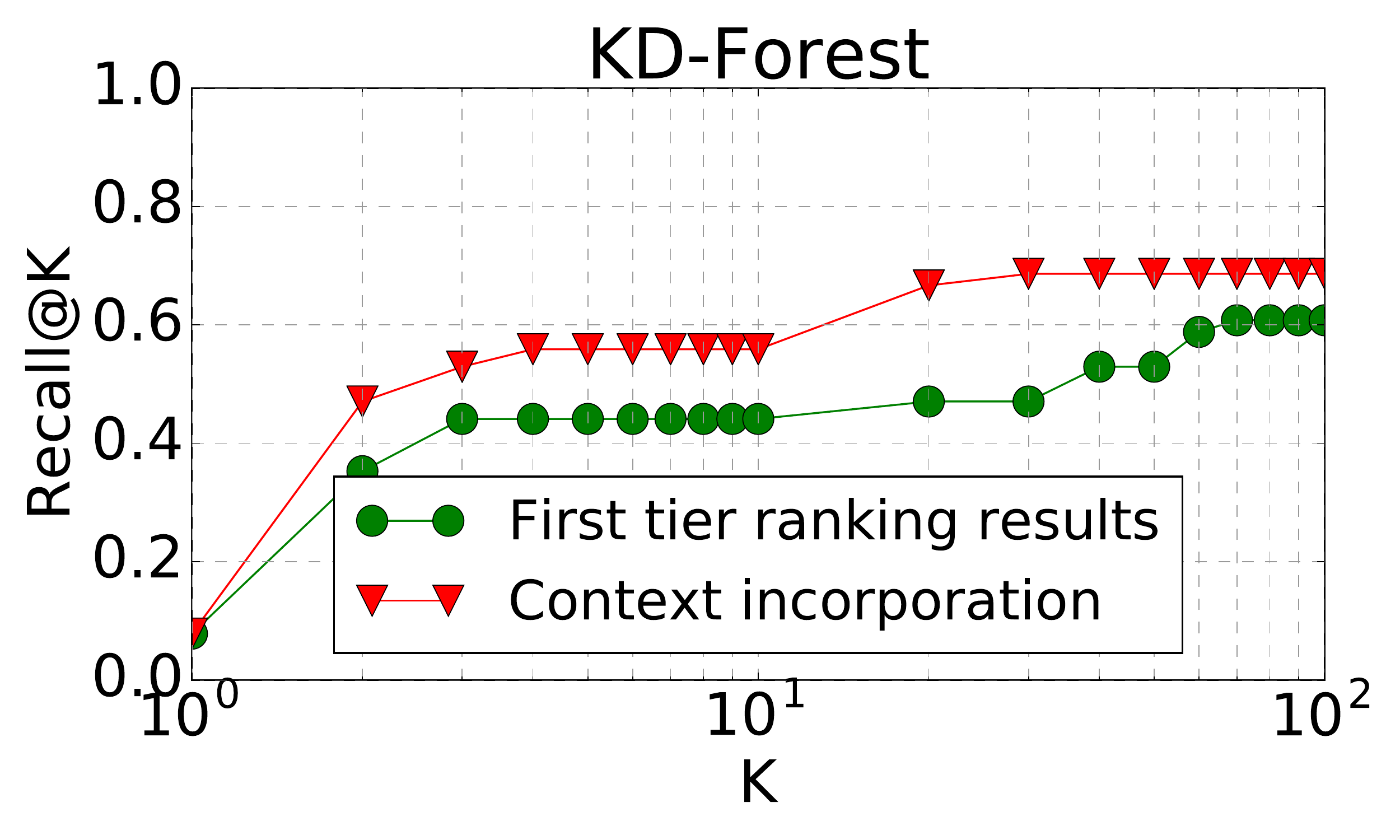}
	\\
	\includegraphics[width=0.48\linewidth]{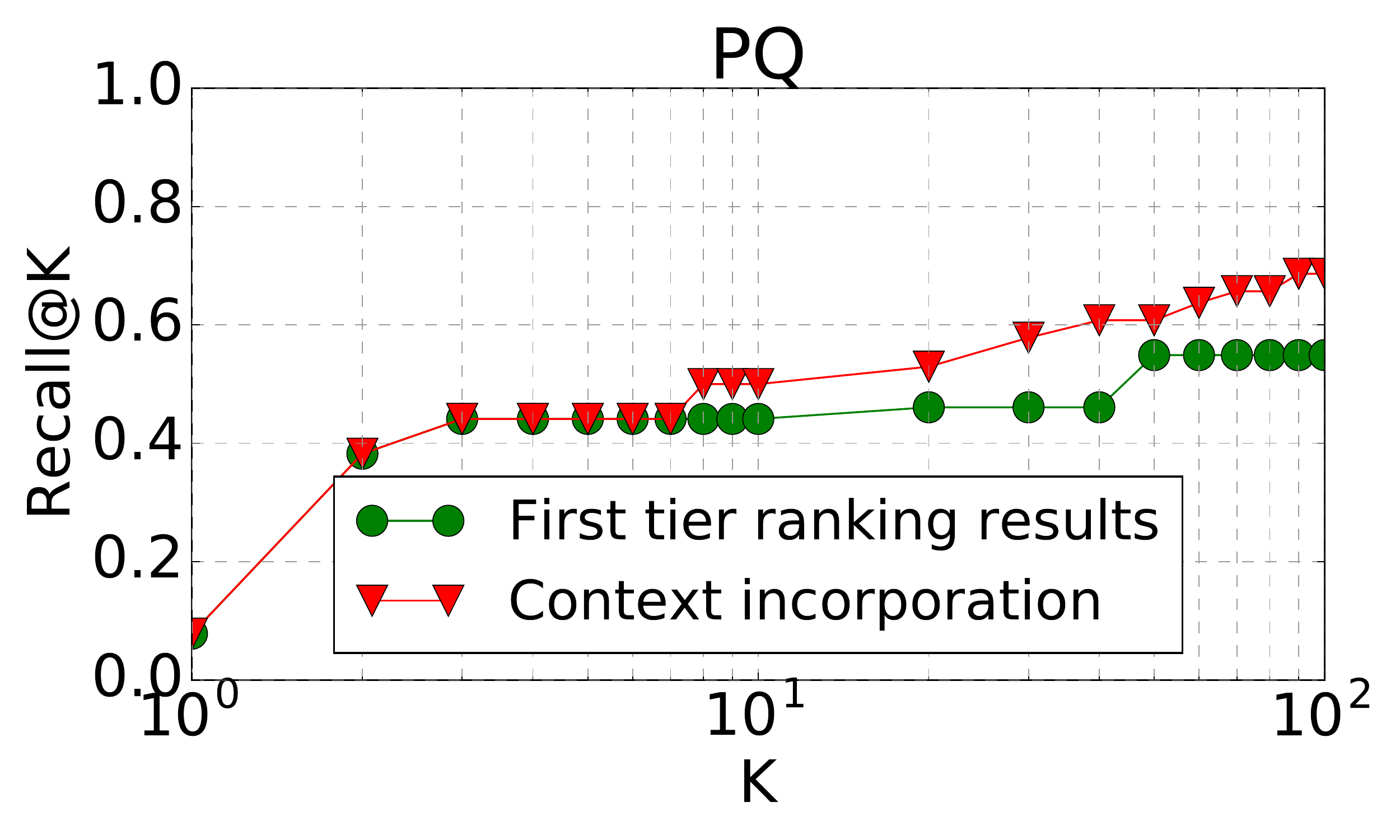}
	\includegraphics[width=0.48\linewidth]{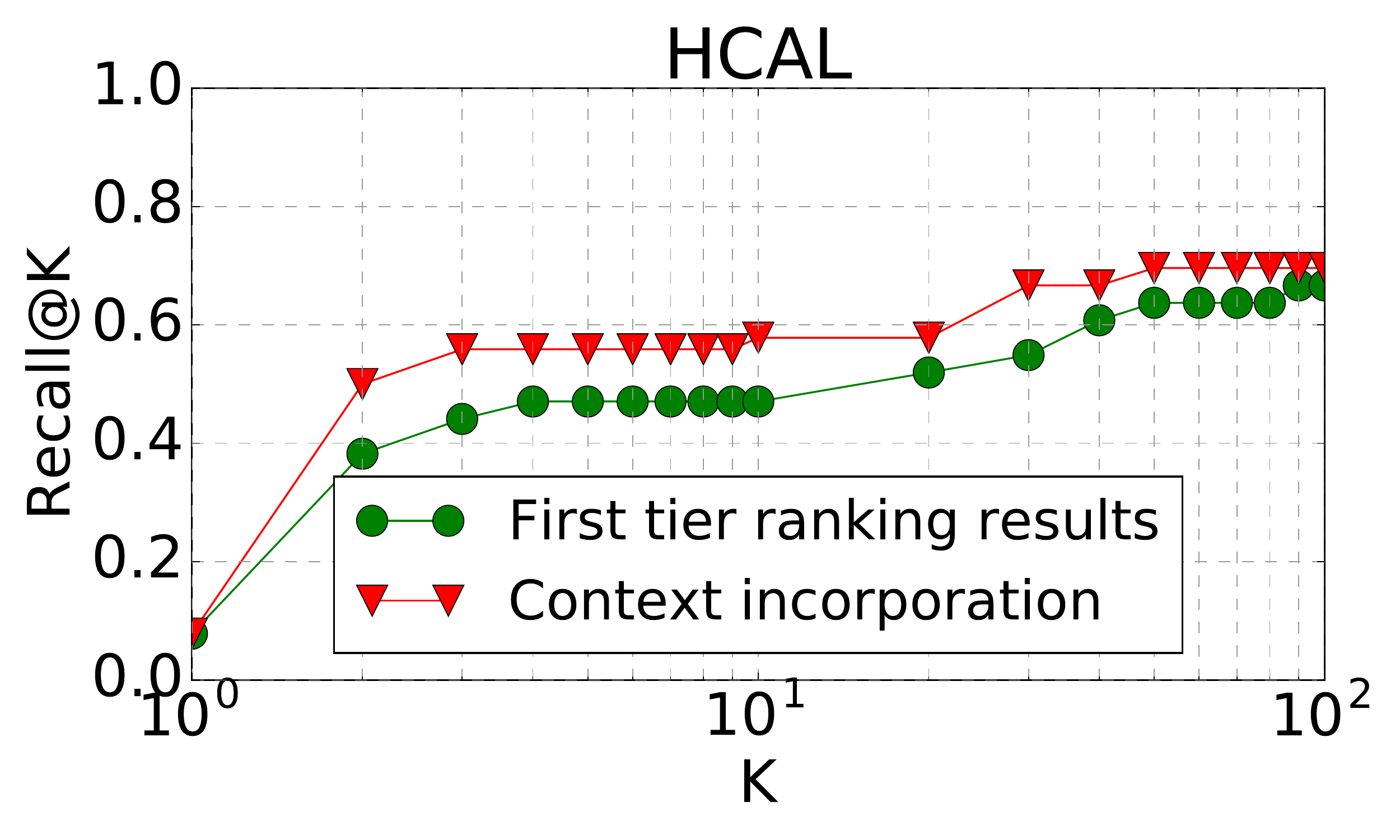}
	\caption{First- and second-tier results for the NC2017 dataset in terms of Recall@k. The context incorporation is important regardless of the  used indexing technique.\label{fig:indexing_compararison_2017}}	
	\end{center}	
\end{figure}

\begin{table}[t]
	\begin{center}
	\begin{footnotesize}	
	\setlength{\tabcolsep}{0.5em}
	\renewcommand{\arraystretch}{0.7}
	\caption{Runtime (in seconds) and memory usage (GB), per query, in the first tier, for different indexing techniques in the NC2017 and NC2017 + World1M datasets. KD-Forest comprises two trees. *~denotes the method did not scale.}
	\label{tab:indexing_method}
	\begin{tabular}{p{2.5cm}p{1.2cm}p{1.5cm}p{1.1cm}p{1.cm}}
		\topline
		\headcol \textbf{Method} & \textbf{KD-Tree} & \textbf{KD-Forest} & \textbf{PQ} & \textbf{HCAL} \\
		\midline
		\textbf{Runtime} & $0.69$ s& $0.72$ s& $13.96$ s& $0.85$ s\\
		\rowcol \textbf{Memory} & $1.48$ GB & $10.69$ GB & $0.02$ GB & $5.38$ GB \\
\hline
		\textbf{Runtime (World1M)} & $8.8$ s& $7.61$ s& $*$ & $*$ \\
		\rowcol \textbf{Memory (World1M)} & $34.99$ GB & $66.42$ GB & $*$  & $*$  \\
		\bottomlinec
	\end{tabular}
	\end{footnotesize}	
	\end{center}	
\end{table}

\vspace*{0.1cm}
\noindent
\textbf{Context Incorporation and Ranking Aggregation.} In this section, we evaluate the proposed approach to improve ranking results for donor images. 
Fig.~\ref{fig:indexing_compararison_2017} shows the performance results in terms of recall at the top-$k$ retrieved images, considering the retrieval of donor images in the first and second tiers of the proposed method. Although not shown here, the performance for retrieving the host image is always above 95\% as it shares much content with $q$. The challenge in provenance filtering is in retrieving the donors.

\vspace*{0.1cm}
\noindent
\textbf{Large-scale Image Retrieval.} We now evaluate the proposed approach, considering a more challenging scenario, in which we embed the NC2016 and NC2017 datasets into one million images, hereinafter referred to as World1M dataset. The World1M dataset contains several images that are semantically similar to the images that compose both datasets. Table~\ref{tab:large_scale_experiments} shows the obtained results in this experiment. There is a gain of about $7\%$ when retrieving donors for  NC2016 when we compare the obtained results in the first and second tiers. The results for NC2017 are slightly lower given that the composite images in this dataset are more difficult, more photorealistic and smaller with respect to the whole image, which also impacts the context incorporation, second tier (first- and second-tier results remain equal for this case). A future work consists of improving the context incorporation mask to better capture small donors such as those present in NC2017. 


\begin{table}[t]
	\begin{center}
	\begin{footnotesize}
	\setlength{\tabcolsep}{0.4em} 
	\renewcommand{\arraystretch}{0.7}
	\caption{Performance results for NC2016 and NC2017 datasets embedded in one million images and KD-Forest (2 trees). Bold highlights improvements in the second tier.\label{tab:wds}}
	\label{tab:large_scale_experiments}
	\begin{tabular}{p{2.8cm}ccc}
		\topline
		\headcol \multicolumn{1}{c}{\textbf{Dataset}}		& \textbf{Type} & \textbf{Tier} & \textbf{Recall@10} \\
		\midline
		
		& 							& 1st 	& $99.65\%$ \\
		\multirow{-2}{*}{\textbf{NC2016 + World1M}}			& \multirow{-2}{*}{{Host}} 	& 2nd 	& $\textbf{100.00}\%$ \\
		\cline{2-4}\rowcol									&  							& 1st 	& $63.00\%$ \\
		\rowcol\multirow{-2}{*}{\textbf{NC2016 + World1M}}	& \multirow{-2}{*}{{Donor}}	& 2nd 	& $\textbf{67.71}\%$ \\
		\hline

		& 							& 1st 	& $88.24\%$ \\
		\multirow{-2}{*}{\textbf{NC2017 + World1M}}			& \multirow{-2}{*}{{Host}} 	& 2nd 	& $88.24\%$ \\
		\cline{2-4}\rowcol									&  							& 1st 	& $25.49\%$ \\
		\rowcol\multirow{-2}{*}{\textbf{NC2017 + World1M}}	& \multirow{-2}{*}{{Donor}}	& 2nd 	& $25.49\%$ \\
		\bottomlinec
	\end{tabular}
	\end{footnotesize}
	\end{center}	
\end{table}

\begin{figure}[t]
	\begin{center}
	\includegraphics[width=0.8\linewidth]{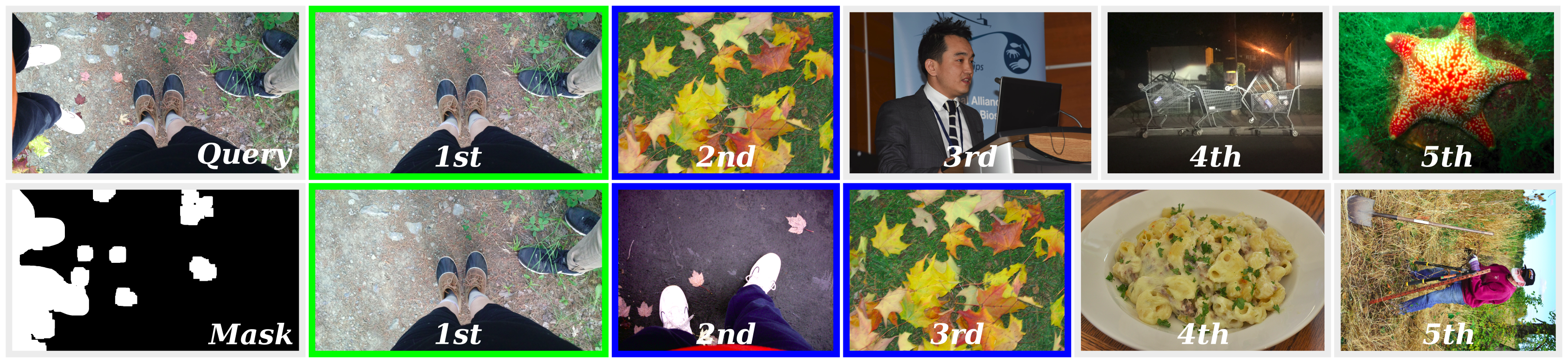}
	\vspace{0.3cm}
	\includegraphics[width=0.8\linewidth]{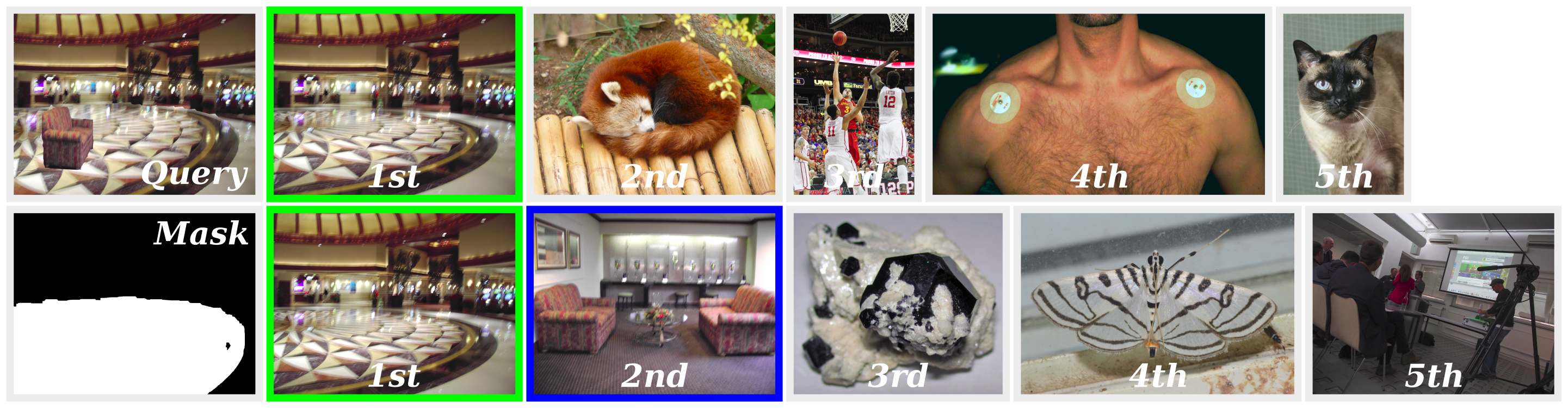}
	\caption{Queries and results for KD-Forest + 2 trees. The first and third rows refer to the first tier results while the second and fourth refer to the second tier. The green border denotes the matched host while the blue ones denote donors. The search in the second tier allows the retrieval of donors that were not present in the first tier.\label{fig:qualitative}}	
	\end{center}	
\end{figure}

\vspace*{0.1cm}
\noindent
\textbf{Qualitative Analysis.} Fig.~\ref{fig:qualitative} shows the results of two queries for KD-Forests with two trees in the first and second tiers. 

\section{Conclusions}
\label{sec:conclusions}
In this paper, we introduced a first method for provenance filtering designed to improve retrieval of donor images in composite images. Reliable provenance filtering is highly useful for selecting the most promising candidates for more complex analyzes in the multimedia phylogeny pipeline such as graph construction and inference of directionality of donors and descendants. 
The challenge in this problem is the retrieval of small objects considering a large image gallery.

By incorporating the context of the top results with respect to the query itself, we can improve the retrieval results and better find possible donors of a given composite (forged) query $q$. Experiments with different indexing techniques have also shown that KD-forests seem to be the most effective but not the most efficient. KD-trees, on the other hand, are more efficient but less effective. In our experiments, PQ did not perform well for large galleries.

Future research efforts will focus on better characterizing small forged regions, incorporating forgery detectors in the process of context analysis and also consider bringing the user into the loop with relevance feedback methods.

\clearpage
\bibliographystyle{IEEEbib}
\bibliography{references}

\begin{thebibliography}{10}

\bibitem{Dias_2013}
Zanoni Dias, Siome Goldenstein, and Anderson Rocha,
\newblock ``Toward image phylogeny forests: Automatically recovering
  semantically similar image relationships,''
\newblock {\em Forensic science international}, vol. 231, no. 1, pp. 178--189,
  2013.

\bibitem{Nimble_2016}
National~Institute of~Standards and Technology (NIST),
\newblock ``The 2016 nimble challenge evaluation dataset,''
  https://www.nist.gov/itl/iad/mig/nimble-challenge-2017-evaluation, Jan. 2016.

\bibitem{Dias_2012}
Z.~Dias, A.~Rocha, and S.~Goldenstein,
\newblock ``Image phylogeny by minimal spanning trees,''
\newblock {\em {IEEE} Transactions on Information Forensics and Security
  (TIFS)}, vol. 7, no. 2, pp. 774--788, April 2012.

\bibitem{Keyes_2004}
Ralph Keyes,
\newblock {\em The post-truth era: Dishonesty and deception in contemporary
  life},
\newblock Macmillan, 2004.

\bibitem{Mahler_2016}
Jonathan Mahler,
\newblock ``The problem with �self-investigation� in a post-truth era,''
\newblock {\em The New York Times Magazine}, January 1st, 2017,
\newblock Available online at \url{http://tinyurl.com/juufufc}.

\bibitem{Schulten_2017}
Katherine Schulten and Amanda~Christy Brown,
\newblock ``Evaluating sources in a `post-truth' world: Ideas for teaching and
  learning about fake news,''
\newblock {\em The New York Times}, January 19th, 2017,
\newblock Available online at \url{http://tinyurl.com/h3w7rp8}.

\bibitem{Rocha:CSUR:2011}
A.~Rocha, W.~Scheirer, T.~E. Boult, and S.~Goldenstein,
\newblock ``{Vision of the Unseen: Current Trends and Challenges in Digital
  Image and Video Forensics},''
\newblock {\em ACM Computing Surveys (CSUR)}, vol. 43, pp. 1--42, 2011.

\bibitem{Oliveira_2016}
Alberto~A de~Oliveira, Pasquale Ferrara, Alessia De~Rosa, Alessandro Piva,
  Mauro Barni, Siome Goldenstein, Zanoni Dias, and Anderson Rocha,
\newblock ``Multiple parenting phylogeny relationships in digital images,''
\newblock {\em IEEE Transactions on Information Forensics and Security}, vol.
  11, no. 2, pp. 328--343, 2016.

\bibitem{Ke_2004}
Yan Ke, Rahul Sukthankar, and Larry Huston,
\newblock ``Efficient near-duplicate detection and sub-image retrieval,''
\newblock in {\em ACM Intl. Conference on Multimedia}, 2004, pp. 869--876.

\bibitem{Zhou:MM:2010}
Wengang Zhou, Yijuan Lu, Houqiang Li, Yibing Song, and Qi~Tian,
\newblock ``Spatial coding for large scale partial-duplicate web image
  search,''
\newblock in {\em {ACM} Int. Conference on Multimedia}, New York, NY, USA,
  2010, MM '10, pp. 511--520, ACM.

\bibitem{Tang:ICIP:2015}
S.~Tang, H.~Chen, K.~Lv, and Y.~D. Zhang,
\newblock ``Large visual words for large scale image classification,''
\newblock in {\em {IEEE} Int. Conference on Image Processing (ICIP)}, Sept
  2015, pp. 1170--1174.

\bibitem{Yuan:ICIP:2015}
J.~Yuan and X.~Liu,
\newblock ``Product tree quantization for approximate nearest neighbor
  search,''
\newblock in {\em {IEEE} Int. Conference on Image Processing (ICIP)}, Sept
  2015, pp. 2035--2039.

\bibitem{Zeng:ICIP:2016}
K.~H. Zeng, Y.~C. Lin, A.~Farhadi, and M.~Sun,
\newblock ``Semantic highlight retrieval,''
\newblock in {\em {IEEE} Int. Conference on Image Processing (ICIP)}, Sept
  2016, pp. 3359--3363.

\bibitem{Dong:ICMR:2012}
Wei Dong, Zhe Wang, Moses Charikar, and Kai Li,
\newblock ``High-confidence near-duplicate image detection,''
\newblock in {\em {ACM} Int. Conference on Multimedia Retrieval}, New York, NY,
  USA, 2012, pp. 1:1--1:8, ACM.

\bibitem{Datta_2008}
Ritendra Datta, Dhiraj Joshi, Jia Li, and James~Z Wang,
\newblock ``Image retrieval: Ideas, influences, and trends of the new age,''
\newblock {\em ACM Computing Surveys (CSUR)}, vol. 40, no. 2, pp. 5, 2008.

\bibitem{Deselaers_2009}
Thomas Deselaers, Tobias Gass, Philippe Dreuw, and Hermann Ney,
\newblock ``Jointly optimising relevance and diversity in image retrieval,''
\newblock in {\em {ACM} Int. Conference on Multimedia Retrieval}. ACM, 2009,
  p.~39.

\bibitem{Dias_2011}
Zanoni Dias, Anderson Rocha, and Siome Goldenstein,
\newblock ``Video phylogeny: Recovering near-duplicate video relationships,''
\newblock in {\em {IEEE} Int. Workshop on Information Forensics and Security
  (WIFS)}. IEEE, 2011, pp. 1--6.

\bibitem{Lameri_2014}
Silvia Lameri, Paolo Bestagini, Ambra Melloni, Simone Milani, Anderson Rocha,
  Marco Tagliasacchi, and Stefano Tubaro,
\newblock ``Who is my parent? reconstructing video sequences from partially
  matching shots,''
\newblock in {\em {IEEE} Int. Conference on Image Processing (ICIP)}. IEEE,
  2014, pp. 5342--5346.

\bibitem{Nucci_2013}
Matteo Nucci, Marco Tagliasacchi, and Stefano Tubaro,
\newblock ``A phylogenetic analysis of near-duplicate audio tracks,''
\newblock in {\em {IEEE} Int. Workshop on Multimedia Signal Processing (MMSP)}.
  IEEE, 2013, pp. 099--104.

\bibitem{Andrews_2012}
Nicholas Andrews, Jason Eisner, and Mark Dredze,
\newblock ``Name phylogeny: A generative model of string variation,''
\newblock in {\em Intl. Conference on Empirical Methods in Natural Language
  Processing and Computational Natural Language Learning}. Association for
  Computational Linguistics, 2012, pp. 344--355.

\bibitem{Bay:CVIU:2008}
Herbert Bay, Andreas Ess, Tinne Tuytelaars, and Luc Van~Gool,
\newblock ``Speeded-up robust features (surf),''
\newblock {\em Comput. Vis. Image Underst.}, vol. 110, no. 3, pp. 346--359,
  June 2008.

\bibitem{Lowe_1999}
David~G Lowe,
\newblock ``Object recognition from local scale-invariant features,''
\newblock in {\em {IEEE} Int. Conference on Computer Vision and Pattern
  Recognition (CVPR)}. Ieee, 1999, vol.~2, pp. 1150--1157.

\bibitem{Zitova_2003}
Barbara Zitova and Jan Flusser,
\newblock ``Image registration methods: a survey,''
\newblock {\em Image and vision computing}, vol. 21, no. 11, pp. 977--1000,
  2003.

\bibitem{Goodfellow_2016}
Ian Goodfellow, Yoshua Bengio, and Aaron Courville,
\newblock {\em Deep learning},
\newblock MIT Press, 2016.

\bibitem{Arya_1998}
Sunil Arya, David~M. Mount, Nathan~S. Netanyahu, Ruth Silverman, and Angela~Y.
  Wu,
\newblock ``An optimal algorithm for approximate nearest neighbor searching
  fixed dimensions,''
\newblock {\em Journal of ACM}, vol. 45, no. 6, pp. 891--923, Nov. 1998.

\bibitem{Bentley:COMMUN:1975}
Jon~Louis Bentley,
\newblock ``Multidimensional binary search trees used for associative
  searching,''
\newblock {\em Commun. ACM}, vol. 18, no. 9, pp. 509--517, Sept. 1975.

\bibitem{Steinbach:KDD:2000}
Michael Steinbach, George Karypis, and Vipin Kumar,
\newblock ``A comparison of document clustering techniques,''
\newblock in {\em In KDD Workshop on Text Mining}, 2000.

\bibitem{Jegou:TPAMI:2011}
H.~Jegou, M.~Douze, and C.~Schmid,
\newblock ``Product quantization for nearest neighbor search,''
\newblock {\em {IEEE} Transactions on Pattern Analysis and Machine Intelligence
  (TPAMI)}, vol. 33, no. 1, pp. 117--128, Jan 2011.

\end{thebibliography}

\end{document}